\documentclass[11pt]{article}
\usepackage{times}
\usepackage{wrapfig}
\usepackage[margin=1in]{geometry}
\usepackage{graphicx}
\usepackage[compact]{titlesec}
\usepackage{color}
\usepackage[numbers]{natbib}
\usepackage{authblk}

\newcommand{\Fig}[1]{Fig.~\ref{#1}}

\begin{document}

\title{\vspace{-1.0cm}\LARGE Cloud No Longer a Silver Bullet, Edge to the Rescue}
\author[1]{Yuhao Zhu}
\author[2]{Gu-Yeon Wei}
\author[2]{David Brooks}
\affil[1]{The University of Texas at Austin}
\affil[2]{Harvard University}
\affil[ ]{\textit{yzhu@utexas.edu}, \textit{\{guyeon, dbrooks\}@eecs.harvard.edu}}

\date{}
\maketitle

\section{Introduction}

The recent surge of cognitive computing presents unprecedented challenges for future computing systems. Siri, a representative cognitive service, has to serve 2 billion requests per week~\citep{siri}. As compared to traditional cloud service requests such as Web search, Siri's voice-based requests not only demand much higher bandwidth, but also require at least two-orders of magnitude higher computation capability~\citep{sirius}. As the number of connected devices keeps increasing, even more data will be generated, which in turn requires enormous computation capability to turn raw data into useful insights. The big question is: how should cognitive services be deployed under such high computation and data pressure?






\begin{figure}[t]
\centering
\vspace*{-5pt}
\includegraphics[trim=0 0 0 0, clip, width=.6\columnwidth]{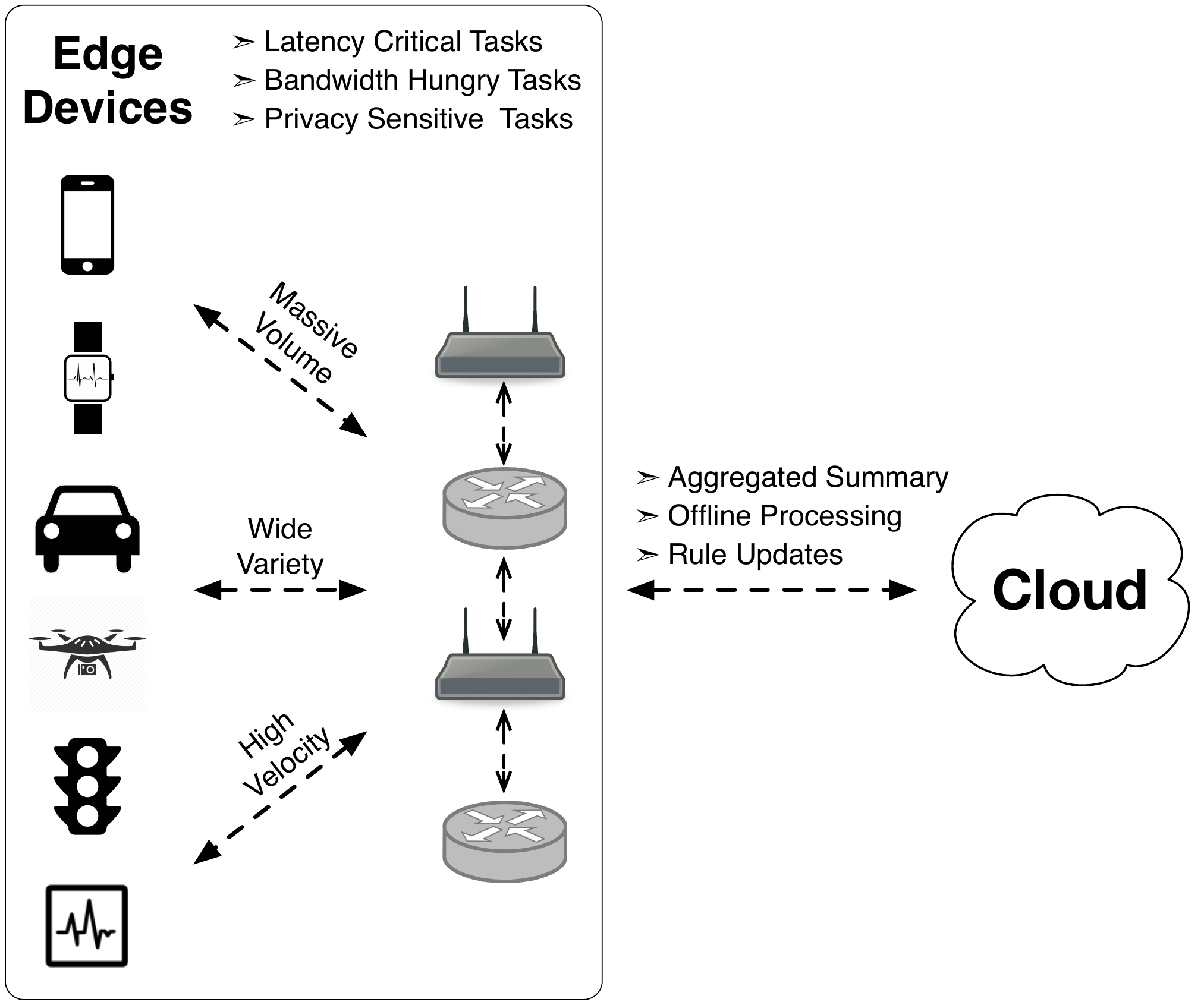}
\caption{The cognitive edge computing infrastructure. Edge devices will handle latency critical, bandwidth hungry, and privacy sensitive tasks while the cloud is used for less-critical tasks such as aggregated summary.}
\label{fig:cogarch}
\vspace*{-5pt}
\end{figure}

This paper takes the position that, while cognitive computing today relies heavily on the cloud, \textit{we will soon see a paradigm shift where cognitive computing primarily happens on network edges}. \Fig{fig:cogarch} illustrates this vision. 
Today's cognitive applications must transfer data 
to a centralized cloud from edge devices, including both end-user devices such as wearables and sensors, as well as network gateways and routers that are just one hop to end-users. We posit that in tomorrow's cognitive computing, edge devices will take over the majority of computations and communicate with each other as a swarm to collectively accomplish tasks.


The shift toward edge devices is fundamentally propelled both by technological constraints in data centers and wireless network infrastructures, as well as practical considerations such as privacy and safety. The remainder of this paper lays out our view of how these constraints will impact future cognitive computing.

Bringing cognitive computing to edge devices opens up several new opportunities and challenges, some of which demand new solutions and some of which require us to revisit entrenched techniques in light of new technologies. We close the paper with a call to action for future research.

\section{Technological Constraints}

Technological constraints of the data center and network infrastructures suggest that deploying cognitive computing tasks (e.g., machine learning model training and inference) at the edge will be more attractive than running exclusively in the cloud.

\paragraph{Data Center Infrastructures}

Cloud computing infrastructure alone is simply not going to be able to provide sufficient computation horsepower within a reasonable energy envelope to carry out cognitive tasks for all connected devices. In contrast, edge devices are poised to provide massive aggregated computation capability in an energy-efficient manner because they are more energy proportional.

Let's again consider Siri as an example. While Siri today is primarily used in smartphones as a voice recognition system, it really represents an emerging class of cognitive service called intelligent personal assistants (IPA). IPAs are rapidly penetrating the consumer market and are projected to soon be ubiquitous---available on other connected devices such as smart home hubs and self-driving cars. Therefore, Siri is a good candidate for miniaturing cognitive computing.

Siri today has to serve 2 billion Siri requests per week~\citep{siri}. This number will increase by as much as one order of magnitude as the number of connected devices reaches 50 billion by 2020~\citep{iotdevices}. At the same time, the computational demand of a Siri-like service is at least two orders of magnitude higher than traditional cloud services such as Web search~\citep{sirius}. This would mean that, even when considering Siri-like cognitive services alone, the total datacenter computation horsepower would have to be three orders of magnitude higher by 2020 than what is available today.

To make up for this huge computation horsepower gap, data centers will have to burn more energy. As the underlying technology scaling that drove the information age comes to an eventual end without practical CMOS alternatives, the energy consumption of the cloud will be prohibitive. Just to put the situation into perspective, data centers today already consume 1.8\% of the total energy consumption in U.S.~\citep{datacenterenergy}. This means that even if we are able to make up for the computation horsepower by scaling out more server processors, our planet simply will not be able to produce enough energy to sustain them!

Pushing cognitive computation to edge devices, in contrast, will allow us to get much higher aggregated compute capability from massive connected edge devices. More importantly, edge devices are more \textit{energy proportional} than data centers due to their low static power designs, resulting in more energy-efficient systems overall~\citep{barroso2007case}. Additionally, hardware designers are also more conscious about energy in designing edge devices as end-users are sensitive to energy consumption for mobile devices.

\paragraph{Wireless Network Infrastructure}

Cognitive computing is undoubtedly associated with ``big data'' due to data-hungry machine learning algorithms. Performing cognitive tasks in the cloud entails massive data transfer between edge devices and the cloud. Unfortunately, the capacity, energy, and latency constraints of (wireless) network infrastructure suggest that massive data transfer is ineffective and often infeasible.

First, the amount of data that connected devices will generate is mind boggling. Cisco projects that mobile data traffic will exceed 30 Exabytes ($10^{18}$) per month by 2020~\citep{ciscomobiledata}, greatly exceeding the total wireless network capacity. Furthermore, data transfer incurs monetary cost, a non-negligible source of expense in many emerging markets where the cost of data could amount to half of a family's income.


Second, the energy overhead of communication is much higher than that of computation. The energy per bit transfer in wireless networks (WiFi and cellular) is on the order of micro Joules ($10^{-6}$)~\citep{huang2012close} while the energy consumption for a 64-bit double precision operation is only on the order of pico Joules ($10^{-12}$)~\citep{dallychallenges}.

Third, many applications and usage scenarios impose tight real-time constraints. Under such circumstances, communication latency is detrimental to application quality-of-service and user quality-of-experience.

Pushing cognitive computing to edge devices could effectively meet the data demand because the majority of data exchanges are limited to within an edge network, effectively filtering or funnelling the data exchange between the edge and the cloud. Each edge network forms a small cell, within which data exchange can leverage the millimeter wave technology that provides extremely high bandwidth, low latency, and low energy connections for devices within a short range~\citep{zhuang2013future}.

\section{Practical Considerations}

Apart from the technological constraints, practical considerations such as privacy and safety also make edge devices more attractive for cognitive computing.

\paragraph{Privacy}

Cognitive computing relies on gathering user-specific data, which often raises serious consumer privacy concerns. For instance, in the wake of spying fears among users Samsung recently had to change their SmartTV privacy policy to explicitly clarify that they do not monitor living room conversations~\citep{samsungsmarttvprivacy}.

In addressing the privacy concern, the White House released the Consumer Privacy Bill of Rights~\citep{pbor}, which outlines the principle of \textit{focused collection} and \textit{data minimization} in gathering consumer data. Focused collection refers to the concept that consumers have a right to reasonable limits on the personal data that companies collect and retain. Data minimization refers to the concept that companies should limit the data they collect and retain, and dispose of it once they no longer need it.

Performing cognitive computing on edge devices in effect practices the focused collection and data minimization principles. If a user's private data is only used for training machine learning models or deriving inferences, there is no reason for the cloud to store or even have access to them.

Companies have already started moving machine learning tasks to edge devices to preserve user privacy. Inference is more amenable to edge devices as it is less computationally intensive. Training, on the other hand, is much more computationally intensive and thus more difficult. One particularly promising approach is \textit{federated learning} where each edge device trains a partial model based on its local dataset and updates to the cloud, which then coordinates all of the partial models to synthesize a global model~\cite{federatedlearning}. Throughout the training process, only model data is communicated while user data remains local.

\paragraph{Safety}

Similar to the tail latency issue in traditional data centers, cognitive computing suffers from the ``tail accuracy'' issue where a small fraction of service requests exhibit poor accuracy due to the stochastic nature of machine learning. In sensitive scenarios such as self-driving cars, even one single mis-prediction could lead to unsafe outcomes.

Edge devices are uniquely equipped to tame the tail accuracy by specializing machine learning models to individuals rather than building generic ones. Specifically, edge devices have direct access to user-specific inference instances, which can be used in-situ to iteratively re-train the model and continually improve accuracy. For instance, an IPA service can best accommodate one's accent by incrementally building a personalized model through interactions with that particular user. Such user-specific data is crucial to improving worst-case error but is different from publicly available proxy datasets used to train the initial model.

\section{Challenges and Opportunities: A Call to Action}



We outline several open research opportunities for computer architects to better connect to the rest of computing society in deploying training and inference engines on the edge. While this is hardly a complete list, we hope to foster discussions that lead to concrete actionable items.

First, we must better understand the characteristics of the cloud, the edge, and the network infrastructures in order to design end-to-end systems that are future-prone. A first-order analytical model that captures the scaling factors of the cloud, edge, and network as well as application requirements will be a key first step.

Second, hardware acceleration on edge devices will be crucial. While there is plenty of work in accelerating inference engines, the next phase of research should also focus on model training. The key, however, is to co-design the training algorithms with the resource-constrained edge devices. Federated learning demonstrates a promising first step.

Last but not least, the privacy policy of a cognitive system should be co-designed with the hardware system to enable flexible trade-offs between privacy, performance, and energy-efficiency~\citep{liu2010tradeoff, zhang2016privacy}. Edge-only computation ensures the strongest privacy but also poses the highest performance and energy requirement. When and how to make trade-offs between different objectives is an open research question.


\newcommand{\BIBdecl}{\setlength{\itemsep}{.2 pt}}
\bibliographystyle{IEEEtranS}
\bibliography{refs}

\end{document}